\documentclass[aps,prl,twocolumn,showpacs]{revtex4}
\newcommand{\be}{\begin{equation}}
\newcommand{\ee}{\end{equation}}
\newcommand{\ben}{\begin{eqnarray}}
\newcommand{\een}{\end{eqnarray}}
\newcommand{\sech}{ \textrm{sech}}
\usepackage{graphicx,epsfig}
\usepackage{amsfonts}
\usepackage{amssymb}
\begin{document}
\title{Solitons in systems of coupled scalar fields}

\author{D. Bazeia, M.J. dos Santos, R.F. Ribeiro}
\affiliation{Departamento de F\'\i sica, Universidade Federal da Para\'\i ba,
C.P. 5008, 58051-970 Jo\~ao Pessoa PB, Brazil}

\date{March 28, 1995}
\begin{abstract}
We present a method to obtain soliton solutions to relativistic system of
coupled scalar fields. This is done by examining the energy associated to
static field configurations. In this case we derive a set of first-order
differential equations that solve the equations of motion when the energy
saturates its lower bound. To illustrate the general results, we investigate
some systems described by polynomial interactions in the coupled fields.
\end{abstract}
\maketitle

Soliton solutions are known to be of great interest in several branches of
physics. They usually apper as solutions of the equations of motion that
describe a dynamical system containging nonlinear interactions among its
degrees of freedom. The issue here is that the basic principles evoked to
find the physical contents of the system, generally lead to a set of coupled
nonlinear second-order differential equations of motion. Unfortunately,
however, there are no general rules for investigating the presence of
soliton solutions for such a set of differential equations.

In this paper we present a method for investigating the presence of soliton
solutions in relativistic systems of coupled scalar fields. The general system
we are going to consider is important in quantum field theory, and in other
branches of physics. In condensed matter, for instance, it may represent the
continuum version of quasi-one-dimensional structures that require at least
two degrees os freedom to be described.

To be explicit, let us consider a general field theory decribing a pair of
interacting real scalar fields in bidimensional spacetime. This system is
governed by the following Lagrangian density.
\begin{equation}
{\mathcal L}=\frac12\partial_\alpha\phi\partial^\alpha\phi+
\frac12\partial_\alpha\chi\partial^\alpha\chi-U(\phi,\chi), 
\label{1}
\end{equation}
where $U(\phi,\chi)$ is the potential, which identifies the particular system
one is interested in. We are using natural units and so the time $t=x^0=x_0$
and the space coordinate $x=x^1=-x_{1}$ have the dimension of inverse energy,
and the fields are dimensionless.

The equations of motion that follow from (\ref{1}) are given by
\begin{eqnarray}
\frac{\partial^2\phi}{\partial t^2}-\frac{\partial^2\phi}{\partial x^2}+
\frac{\partial V}{\partial \phi}=0 \label{2}\\
\frac{\partial^2\chi}{\partial t^2}-\frac{\partial^2\chi}{\partial x^2}+
\frac{\partial V}{\partial \chi}=0 \label{3}
\end{eqnarray}
For static field configurations $\phi=\phi(x)$ and $\chi=\chi(x)$ we get
\begin{eqnarray}
\frac{d^2\phi}{dx^2}=\frac{\partial V}{\partial \phi}, \label{4}\\
\frac{d^2\chi}{dx^2}=\frac{\partial V}{\partial \chi}. \label{5}
\end{eqnarray}
The potential $U(\phi,\chi)$ in general is a nonlinear function of the
coupled fields, and so the above equations of motion are coupled nonlinear
second-order ordinary differential equations. In the standard
way \cite{1,2,3,4} of  searching for soliton solutions, one has
to deal with the above equations of motion.

Here, however, we shall follow another route \cite{5}. In this case we
investigate the energy corresponding to the static field configurations.
For the above system, the energy can be written as
\begin{equation}
E=\frac12\int^{\infty}_{-\infty}dx\left[\left(\frac{d\phi}{dx}\right)^2+
\left(\frac{d\chi}{dx}\right)^2+2U(\phi,\chi)\right].\label{6}
\end{equation}
In order to present a general procedure, let us now introduce tow arbitrary
but smooth functions $F=F(\phi,\chi)$ and $G=G(\phi,\chi)$. Then, we work
with the tow first terms in expression (\ref{6}) to rewrite the energy in
the following form,
\begin{equation}
E=E_b+E', \label{7}
\end{equation}
where $E_b$ is given by
\begin{eqnarray}
E_b&=& \left|{\mathcal F} [\phi(\infty),\chi(\infty)]-{\mathcal F}
[\phi(-\infty),\chi(-\infty)]\right| \nonumber \\
&+&\left|{\mathcal G} [\phi(\infty),\chi(\infty)]-{\mathcal G}
[\phi(-\infty),\chi(-\infty)]\right| \label{8}
\end{eqnarray}
with ${\mathcal F}(\phi,\chi)=\phi F(\phi,\chi)$ and
${\mathcal G}(\phi,\chi)=\chi G(\phi,\chi)$, and $E'$ has the form
\begin{eqnarray}
E'&=&\frac12\int^{\infty}_{-\infty}dx\left\{\left(\frac{d\phi}{dx}+F+
\phi\frac{\partial F}{\partial \phi}+\chi\frac{\partial G}
{\partial\phi}\right)^2\right. \nonumber\\
&+&\left(\frac{d\chi}{dx}+G+\phi\frac{\partial F}
{\partial \chi}+\chi\frac{\partial G}{\partial\chi}\right)^2 \nonumber\\
&+&\left[2U-\left(F+\phi\frac{\partial F}{\partial \phi}+
\chi\frac{\partial G}{\partial\phi}\right)^2\right. \nonumber\\
&-&\left.\left.\left(G+\phi\frac{\partial F}{\partial \chi}+
\chi\frac{\partial G}{\partial\chi}\right)^2\right]\right\}.\label{9}
\end{eqnarray}
From expression (\ref{9}) we recognize that if the potencial has the specific
form
\begin{equation}
U=\frac12\left(F+\phi\frac{\partial F}{\partial \phi}+
\chi\frac{\partial G}{\partial\phi}\right)^2+\frac12
\left(G+\phi\frac{\partial F}{\partial \chi}+
\chi\frac{\partial G}{\partial\chi}\right)^2,	\label{10}
\end{equation}
the energy is saturated to its bound $E_b$ given in (\ref{8}) for static 
field configurations satisfying the following set of first-order equations
\begin{eqnarray}
\frac{d\phi}{dx}+F+\phi\frac{\partial F}{\partial \phi}+
\chi\frac{\partial G}{\partial\phi}=0. \label{11}\\
\frac{d\chi}{dx}+G+\phi\frac{\partial F}{\partial \chi}+
\chi\frac{\partial G}{\partial\chi}=0, \label{12}
\end{eqnarray}
We now work with Eqs. (\ref{11}) and (\ref{12}) to verify that any pair
$(\phi,\chi)$ that solves this set of first-order equations also solves
the second-order equations of motion (\ref{4}) and (\ref{5}) when the
potencial is given by expression (\ref{10}). This procedure then changes
the task of solving sets of second-order differential equations to the
task of solving sets of first-order differential equations. The price
one pays is that this is not always possible, since the potencial must
have the specific form given by expression (\ref{10}). When this is the
case, however, we search for the minimum energy configurations that will
certainly play an important physical role, and whose energy one calculates
immediately via expression (\ref{8}).

The above precedure is completely general, and can be generalized to
systems of three or more fields straightforwardly. However, to investigate
the presence of topological solitons we impose the restrictions
\begin{eqnarray}
U(\phi,\chi)=U(-\phi,\chi), \nonumber\\
U(\phi,\chi)=U(\phi,-\chi). \label{13}
\end{eqnarray}
Here we are guided by the fact that parity symmetry is necessary (but not
sufficient) to give rise to topological solitons. Now, from (\ref{10}) and
(\ref{13}) one sees that the functions $F$ and $G$ must have opposite parity
in each of the fields: if $F(\phi,\chi)=\pm F(-\phi,\chi)$, then
$G(\phi,\chi)=\mp G(-\phi,\chi)$, and if $F(\phi,\chi)=\pm F(\phi,-\chi)$,
then $G(\phi,\chi)=\mp G(\phi,-\chi)$.

To investigate particular systems, we must specify the functions $F$ and $G$.
We do this by using dimensionless (a,b,\ldots) {\it real and positive} and
dimensional $(\lambda,\mu,\ldots)$ {\it real} parameters - dimensional
parameters have the dimension of energy. First we note that if one chooses
$F=F(\phi)$ and $G=G(\chi)$, the two scalar fields decouple. As a simple
illustration, let us consider $F=F(\phi)$ and $G=0$. Here we get
\begin{eqnarray}
U(\phi)=\frac12\left(F+\phi\frac{dF}{d\phi}\right)^2,  \label{14}\\
\frac{d\phi}{dx}+F+\phi\frac{dF}{d\phi}=0.  \label{15}
\end{eqnarray}
We now choose $F$ in the form
\begin{equation}
F(\phi)=\lambda(\frac13\phi^2-a^2), \label{16}
\end{equation}
to get
\begin{equation}
U(\phi)=\frac12\lambda^2(\phi^2-a^2), \label{17}
\end{equation}
and so the first-order equation becomes
\begin{equation}
\frac{d\phi}{dx}+\lambda(\phi^2-a^2)=0. \label{18}
\end{equation}
This is the $\phi^4$ system. It has the well known \cite{4} solutions
\begin{equation}
\phi(x)=a\tanh[\lambda a(x+\overline{x})]. \label{19}
\end{equation}
Since we are considering $a>0$, the sing of $\lambda$ specifies the particular
solution: $\lambda>0$ for the kink, and $\lambda<0$ for the antikink. The
corresponding energy is given by 
\begin{equation}
E=\frac43|\lambda|a^3. \label{20}
\end{equation}
Another simple illustration is obtained by choosing  $F=F(\phi)$ in the form
\begin{equation}
F(\phi)=\frac12\lambda\phi(\frac12\phi^2-a^2). \label{21}
\end{equation}
In this case we get
\begin{equation}
U(\phi)=\frac12\lambda^2\phi^2(\phi^2-a^2)^2, \label{22}
\end{equation}
and
\begin{equation}
\frac{d\phi}{dx}+\lambda\phi(\phi^2-a^2)=0. \label{23}
\end{equation}
This is the $\phi^6$ system. It has \cite{6} soliton solutions,
which we can write as
\begin{equation}
\phi^2(x)=\frac12a^2\left\{1+
\tanh[\lambda a^2(x+\overline{x})]\right\}.\label{24}
\end{equation}
Note that the sign of $\lambda$ identifies different solutions.
The correponding energy is
\begin{equation}
E=\frac14|\lambda|a^4. \label{25}
\end{equation}
To go to a more general situation we now consider $F=F(\phi)$ and
$G=G(\phi,\chi)$. From the reasoning just below Eqs. (\ref{13}), we note
that $G(\phi,\chi)=-G(\phi,-\chi)$, and that $F(\phi)$ and $G(\phi,\chi)$
must have an opposite parity in $\phi$. Then, if one chooses for $F(\phi)$
the same function we have already considered in (\ref{16}), the simplest
choice for $G(\phi,\chi)$ is given by
\begin{equation}
G(\phi,\chi)=\frac12\mu\phi\chi. \label{26}
\end{equation}
Here the system is specified by potencial
\begin{equation}
U(\phi,\chi)=\frac12\left[\lambda(\phi^2-a^2)+
\frac12\mu\chi^2\right]^2+\frac12\mu^2\phi^2\chi^2. \label{27}
\end{equation}
The corresponding first-order equations that follow from (\ref{11}) and
(\ref{12}) are given by 
\begin{eqnarray}
\frac{d\phi}{dx}&+&\lambda(\phi^2-a^2)+\frac12\mu\chi^2=0,\label{28}\\
\frac{d\chi}{dx}&+&\mu\phi\chi=0. \label{29}
\end{eqnarray}
The above system comprises soliton solutions. To find some of them,
first one set $\phi=0$. This implies that $\chi^2=2(\lambda/\mu)a^2$,
and so no soliton solution is found; note, however, that the parameters
$\lambda$ and $\mu$ must have the same sign. Next, we set $\chi=0$. In
this case, the first-order equations (\ref{28}) and (\ref{29}) reduce to
the single equation (\ref{18}). Here the pair of solutions is given by
\begin{equation}
\phi(x)=a \tanh[\lambda a(x+\overline{x})], \qquad \chi(x)=0. \label{30}
\end{equation}
The corresponding energy is calculated immediately, and give the same value
(\ref{20}) we have already obtaind for the $\phi^4$ system.

To find other soliton solutions we use the trial orbit method introduced by
Rajaraman \cite{2} to choose an orbit. Here, however, we are dealing with
first-order equations and this seems to ease the task of choosing orbits.
For instance, from Eq. (\ref{29}) one sees that orbits like
$\phi^2-a^2=b\chi^2$ and $\phi^2-a^2=b\chi^2+c\chi$ are good ones.
Moreover, the orbit one chooses must be compatible with the first-order
equation (\ref{28}) and (\ref{29}), and this provides a direct verification
of the trial choice one is testing. We follow this reasoning to choose
\begin{equation}
\lambda(\phi^2-a^2)+\frac12\mu\chi^2=\mu(\phi^2-a^2). \label{31}
\end{equation}
In this case Eq. (\ref{28}) changes to
\begin{equation}
\frac{d\phi}{dx}+\mu(\phi^2-a^2)=0. \label{32}
\end{equation}
We solve this equation to get
\begin{equation}
\phi(x)=a \tanh [\mu a(x+\overline{x})], \label{33}
\end{equation}
and so
\begin{equation}
\chi(x)=\pm a [2(\lambda/\mu-1)]^{1/2}\sech[\mu a(x+\overline{x})], \label{34}
\end{equation}
which requires $|\lambda|>|\mu|$. To calculate the energy corresponding to the
above solutions we use expression (\ref{8}): it is independent of $\mu$ and
has the same value we have already calculated in (\ref{20}).

To get another pair of solutions we set
\begin{equation}
\phi^2-a^2=\chi^2\pm2\frac{ab}{\sqrt{b^2-1}}\chi, \label{35}
\end{equation}
in which $b>1$. In this case the first-order Eq. (\ref{29}) changes to
\begin{equation}
\frac{d\chi}{dx}+\mu\left(a^2\pm2\frac{ab}{\sqrt{b^2-1}}\chi+
\chi^2\right)^{1/2} \chi=0, \label{36}
\end{equation}
which can be integrated to give
\begin{equation}
\chi(x)=\mp a\frac{\sqrt{b^2-1}}{b+\cosh[\mu a(x+\overline{x})]},
\label{37}
\end{equation}
and so we get
\begin{equation}
\phi(x)=a\frac{\sinh[\mu a(x+\overline{x})]}
{b+\cosh[\mu a (x+\overline{x})]},\label{38}
\end{equation}
which requires $\mu=2\lambda$. Once again, the energy is given by
expression (\ref{20}), and now it is independent of $b$.

As we have explicitly shown, the above soliton solutions (\ref{30}),
(\ref{33}) and (\ref{34}), and (\ref{37}) and (\ref{38}) have the same
energy. This is so because they belong to the same topological sector.

To investigate another system, let us choose $F(\phi)$ in the form give
in (\ref{21}). In this case the simplest choice for $G(\phi,\chi)$
is given by
\begin{equation}
G(\phi,\chi)=\frac12\mu\phi^2\chi, \label{39}
\end{equation}
and so the system is specified by the potential
\begin{equation}
U(\phi,\chi)=\frac12[\lambda\phi(\phi^2-a^2) +\mu\phi\chi^2]^2+
\frac12\mu^2\phi^4\chi^2. \label{40}
\end{equation}
The corresponding first-order equations that now follow are given by
\begin{eqnarray}
\frac{d\phi}{dx}&+&\lambda\phi(\phi^2-a^2)+\mu\phi\chi^2=0, \label{41}\\
\frac{d\chi}{dx}&+&\mu\phi^2\chi=0. \label{42}
\end{eqnarray}
The above system comprises soliton solutions. To find some of them,
first one sets $\phi=0$. This implies that $\chi$ is constant, and so no
soliton solution is found. Next, we set $\chi=0$, which reduces the above
set of first-order equations to the singles Eq. (\ref{23}). Here the pair
of solutions is given by
\begin{eqnarray}
\phi^2(x)&=&\frac12a^2\{1+\tanh[\lambda a^2(x+\overline{x})]\},\nonumber\\
\chi(x)&=&0. \label{43}
\end{eqnarray}
The correponding energy is calculated immediately, and gives the same
value (\ref{25}) we have already obtained for the $\phi^6$ system.

To find other soliton solutions to (\ref{41}) and (\ref{42}) we follow
the reasoning we have already considered in the former system. Here we set
\begin{equation}
\lambda\chi(\phi^2-a^2)+\mu\phi\chi^2=\mu\phi(\phi^2-a^2). \label{44}
\end{equation}
In this case the first-order equation (\ref{41}) changes to 
\begin{equation}
\frac{d\phi}{dx}+\mu\phi(\phi^2-a^2)=0. \label{45}
\end{equation}
The solutions to this equation are
\begin{equation}
\phi^2(x)=\frac12a^2\{1+\tanh[\mu a^2(x+\overline{x})]\}, \label{46}
\end{equation}
and so we get
\begin{equation}
\chi^2(x)=\frac12a^2(\lambda/\mu-1)\{1-\tanh[\mu a^2(x+\overline{x})]\},
\label{47}
\end{equation}
which requires $\lambda$ and $\mu$ to have the same sign, and
$|\lambda|>|\mu|$. To calculate the energy corresponding to the
above solutions we use expression (\ref{8}): it is independent of
$\mu$ and has the same value we have already calculated in(\ref{25}).

We now investigate one further system. Here we choose the function
$F=F(\phi)$ yet as an odd function in $\phi$, but now we consider
the simplest form
\begin{equation}
F(\phi)=-\frac12\lambda a^2\phi. \label{48}
\end{equation}
In this case the function $G=G(\phi,\chi)$ has to be even in $\phi$
and odd in $\chi$. As a simple choice we write
\begin{equation}
G(\phi,\chi)=\frac12\lambda(\phi^2-a^2)\chi. \label{49}
\end{equation}
In this case the system is specified by the following potential
\begin{equation}
U(\phi,\chi)=\frac12\lambda^2\phi^2(\chi^2-a^2)^2+
\frac12\lambda^2\chi^2(\phi^2-a^2)^2, \label{50}
\end{equation}
and the corresponding first-order equations are given by
\begin{eqnarray}
\frac{d\phi}{dx}+\lambda(\chi^2-a^2)\phi=0, \label{51}\\
\frac{d\chi}{dx}+\lambda(\phi^2-a^2)\chi=0. \label{52}
\end{eqnarray}
We now choose the orbit $\phi^2=\chi^2$, and so the solutions are
\begin{equation}
\phi^2(x)=\chi^2(x)=\frac12a^2
\{1+\tanh[\lambda a^2(x+\overline{x})]\}.\label{53}
\end{equation}
Like in the former system, here the sign of $\lambda$ also identifies
different solutions. To calculate the corresponding energy, we use
expression (\ref{8}) to see that it has the same value we have already
calculated in (\ref{25}).

An issue that now follows concerns investigating the stability of the
static field configurations we have found. To give an explicit calculation,
let us consider solutions (\ref{30}) to investigate the classical or
linear \cite{2} stability in the simplest case. Here we write
\begin{eqnarray}
\phi(t,x)=\phi(x)+g(x)\cos(\varepsilon_1t), \label{54}\\
\chi(t,x)=\chi(x)+h(x)\cos(\varepsilon_2t), \label{55}
\end{eqnarray}
where $\phi(x)$ and $\chi(x)$ are given by (\ref{30}). We now use the
equation of motion (\ref{2}) and (\ref{3}) to get
\begin{eqnarray}
\frac{d^2g}{dz^2}+[w_1^2-4+6\sech^2(z)]g=0, \label{56}\\
\frac{d^2h}{dz^2}+\left[w_2^2-\frac{\mu^2}{\lambda^2}+
\frac\mu\lambda\left(\frac\mu\lambda+1\right)\sech^2(z)\right]h=0. \label{57}
\end{eqnarray}
where $z=\lambda a(x+\overline{x})$ and
$\omega_i^2=\varepsilon_i^2/\lambda^2 a^2$, with $i=1,2.$ Eq. (\ref{56})
is just the equation we get in the $\phi^4$ system. It has \cite{7} tow
bound states, $\omega^2_{1,0}=0$ and $\omega^2_{1,1}=3$. Eq. (\ref{57})
has bound states given by
\begin{equation}
\omega^2_{2,n}=(2\mu/\lambda-n)n, \label{58}
\end{equation}
where $n=0,1,\ldots<\mu/\lambda$. There is no negative energy bound state,
and so the set of solutions (\ref{30}) is classically stable, at every point
in the parameter space. This result is interesting since one knows \cite{8}
that similar static field configurations that solve the system considered in
Ref. \cite{3} are linearly unstable, at least in some region in parameter
space.

For the set os solutions (\ref{43}), a similar investigation shows that it
is also classically stable. The other solutions are harder to investigate,
because the fluctuations become coupled and this greatly complicates the
calculations concerning classical stability. We shall return to this specifc
issue in a separate work \cite{9}. Here we just advance that all the soliton
solutions we have found are classically stable.

To end, we recall that we have investigated the energy corresponding to
static field configurations to offer a method to find soliton solutions
to systems of coupled scalar fields. To illustrate the procedure, we have
dealt with some specific systems, containing the forth and the sixth powers
in the coupled fields. We then used the trial orbit method introduced by
Rajaraman \cite{3} to find soliton solutions. Evidently, the trial orbit
method is not the only route. For instance, some very recent
investigations \cite{10,11} have introduced other methods,
which can perhaps offer more soliton solutions for the system
we have just investigated.

As we have shown, the method seems to work well, at least for polynomial
interactions between the coupled fields. However, further investigations
are desired, mainly the one concerning the generalization to nonpolynomial
interactions, which can be important to model periodic structures described
by tow or more degrees of freedom. This and other related issues are now
under consideration.

We would like to thank interesting discussions with M.A. de Melo Gomes.
DB is grateful to Conselho Nacional de Desenvolvimento Cient\'\i fico e
Tecnol\'ogico, CNPq, Brazil, for partial support. MJdS is thankful to
Coordena\c c\~ao de Aperfei\c coamento de Pessoal do Ensino Superior,
CAPES, Brazil, for a fellowship.

\end{document}